\documentclass[3p]{elsarticle}

\usepackage{hyperref}
\usepackage{color}
\usepackage{graphicx}
\usepackage{dcolumn}
\usepackage{bm}
\usepackage{braket}
\usepackage{dsfont}
\usepackage{color}
\usepackage{float}
\usepackage{textgreek}
\usepackage[LGR,T1]{fontenc}
\usepackage{array}
\usepackage{textcomp}
\usepackage{multirow}

\journal{Journal of \LaTeX\ Templates}

\begin{document}

\begin{frontmatter}

\title{DNP-NMR of surface hydrogen on silicon microparticles}

\author{Daphna Shimon$^a$}
\ead{daphna.shimon@dartmouth.edu}
\author{Kipp J. van Schooten$^a$}
\address[1]{Department of Physics and Astronomy, Dartmouth College, Hanover,
NH 03755, U.S.A.}
\author{Subhradip Paul$^b$}
\address[2]{School of Physics and Astronomy, University of Nottingham, Nottingham NG7 2RD, UK.}
\author{Zaili Peng$^c$}
\author{Susumu Takahashi$^{c,d}$}
\address[3]{Department of Chemistry, University of Southern California, Los Angeles, California 90089, USA}
\address[4]{Department of Physics and Astronomy, University of Southern California, Los Angeles, California 90089, USA}
\author{Walter K\"ockenberger$^b$}
\author{Chandrasekhar Ramanathan$^a$}
\ead{chandrasekhar.ramanathan@dartmouth.edu}

\begin{abstract}
Dynamic nuclear polarization (DNP) enhanced nuclear magnetic resonance (NMR) offers a promising route to studying local atomic environments at the surface of both crystalline and amorphous materials. We take advantage of unpaired electrons due to defects close to the surface of the silicon microparticles to hyperpolarize adjacent $^{1}$H nuclei. At 3.3 T and 4.2 K, we observe the presence of two proton peaks, each with a linewidth on the order of 5 kHz. Echo experiments indicate a homogeneous linewidth of $\sim 150-300$ Hz for both peaks, indicative of a sparse distribution of protons in both environments. The downfield peak at 10 ppm lies within the typical chemical shift range for proton NMR, and was found to be relatively stable over repeated measurements. The upfield peak was found to vary in position between -19 and -37 ppm, well outside the range of typical proton NMR shifts, and indicative of a high-degree of chemical shielding. The upfield peak was also found to vary significantly in intensity across different experimental runs, suggesting a weakly-bound species. These results suggest that the hydrogen is located in two distinct microscopic environments on the surface of these Si particles.
\end{abstract}

\begin{keyword}
NMR, DNP, nuclear magnetic resonance, dynamic nuclear polarization, silicon particles, surface
\end{keyword}

\end{frontmatter}


\section{Introduction}

\noindent Studying surfaces with nuclear magnetic resonance (NMR) has been a goal of the magnetic resonance community for many years because it can reveal the atomic environments of crystalline or amorphous materials. However, the lack of sensitivity makes this notoriously difficult. Dynamic nuclear polarization
(DNP) can dramatically enhance the NMR signal
and has recently been shown to enable surface-NMR.
DNP of surfaces very often involves wetting of the surface of the particles using exogenous radicals dissolved in a solvent.
The radicals are used as the source for DNP enhancement of either the surface of the particles via cross polarization (CP)\cite{RN273} from $^{1}$H nuclei in the solvent to the surface heteronuclei, or directly to the bulk heteronuclei. \cite{RN128,RN119,RN112,RN117,RN114,RN105,RN120,RN111,RN123,RN104,RN103,RN109,RN118,RN126,RN115,RN136} These techniques are most commonly combined with magic angle spinning (MAS) at 100 K for higher resolution solid-state NMR spectra.
 
Several groups have also used endogenous paramagnetic centers inside the studied material as the DNP enhancement source.\cite{RN174,RN333,RN334,RN335,RN294,RN296,RN297,RN303,RN347} With this technique, defects that are intrinsic to the microparticles, or metals, are used as a source of enhancement, thus not altering the surface with solvents or exogenous radicals - potentially avoiding surface degradation or adding nuclei to the sample. Though limited to a smaller range of material systems, this type of DNP has been explored with NV-centers in diamonds,\cite{RN294,RN296,RN297,RN303} as well as in battery materials,\cite{RN334} other doped crystalline materials,\cite{RN333} biological samples, \cite{RN353} and in silicon particles.\cite{RN335,RN347,RN175,RN171,RN174,RN163}   
 
Silicon nano- and microparticles attract much interest in the scientific community because of their bio-compatibility and their varied potential uses. For example, they can be used 
for biomedical magnetic resonance imaging, for batteries, photovoltaics and other uses. \cite{RN171,RN173,RN175,RN199,RN265,RN266,RN267,RN268} However, these particles are prone to degrade via exposure to air and humidity. Water molecules are known to chemisorb onto the surface of silicon, resulting in oxidized surface layer consisting of Si-H and Si-OH groups, through dissociation of water molecules at the sites of dangling bonds \cite{RN166,RN165,RN168,RN343} This oxidation can result in degradation of the silicon surface and affect the electronic and material properties. \cite{RN169,RN170,RN286} It is therefore important to develop tools to study surface structure and chemistry in these systems. 
 
Silicon particles are known to contain localized  unpaired electrons at the interface between the silicon and the silicon-oxide layers.\cite{RN274,RN275,RN276,RN163,RN336,RN337,RN338} At room temperature these Si\textminus H and Si\textminus OH species are able to move around the surface to adjacent unoccupied dangling bonds, if any. These systems also exhibit a significant degree of heterogeneity based on the conditions under which the host materials were grown, the methods by which the powders were prepared and sorted, and the storage conditions. For example the density of dangling bond defects at the Si/SiO$_2$ interface can vary by over 2 orders of magnitude, significantly changing the efficiency of DNP experiments.\cite{RN336,RN337,RN338} The presence of other dopants and defects can also result in DNP variations. \cite{RN175}  This can be a challenge in using DNP-NMR for materials characterization.

Here, we study the surface of silicon microparticles 
by detecting $^{1}$H nuclei adjacent to the surface defects. \cite{RN176,RN177,RN178,RN189,RN190,RN192,RN193,RN194,RN196,RN284,RN269,RN272} 
At 3.3 T and 4.2 K, we observe the presence of two proton peaks each with a linewidth on the order of 5 kHz, indicating two distinct microscopic surface environments. We characterize the NMR lineshapes, the DNP spectra and the characteristic relaxation times of the observed spectra. MW frequency modulation was found to result in a large increase in DNP enhancement for both peaks.

\section{Results and Discussion}

\noindent The DNP experiments were performed at a field of 3.3 T, corresponding to an electron Larmor frequency of 94 GHz and a $^{1}$H Larmor frequency of 142 MHz. The setup was
equipped with a Janis continuous-flow NMR cryostat, enabling experiments at 4.2K. The NMR pulses and detection were controlled with a Bruker Avance AQX spectrometer. The MW irradiation was produced via a previously described millimeter wave source. \cite{RN163,RN164}  

Powdered silicon particles (nominally 1-5\textmu m poly crystalline silicon) were bought from Alfa Aesar, and were used as is. The powder was stored for many years in the original packaging, under ambient conditions before use. Several samples were made from the same batch, listed A-D. Sample A was cooled down to 4 K twice for two sets of experiments, listed as A1 and A2. A2 was measured 17 days after sample A1. For each sample, the powder was tightly packed into a glass capillary and flame sealed to prevent further contact with ambient humidity, but air and ambient humidity were not removed before sealing.

The $^{1}$H signal was collected via solid echo detection, \cite{RN278} \cite{RN55}
with a \textpi /2=0.5 \textmu s excitation (approximately 500 kHz RF power)
and an echo delay of \texttau =200 \textmu s. The echo time was used to minimize contributions from the probe background. No change was observed in the line-widths of the DNP-enhanced peaks using echo times of 30 $\mu$s and 200 $\mu$s (see Supporting Information).
Two step phase cycling was used to subtract the free induction decay signal from the second \textpi /2
pulse. All experiments employed only 2 scans, unless otherwise noted. All experiments began with a train of pulses used to saturate the $^{1}$H nuclei. In all cases the MW irradiation was turned off during
NMR signal acquisition.

\subsection{NMR Lineshapes}

\begin{figure}[H]
\begin{center}
\includegraphics[width=0.7\textwidth]{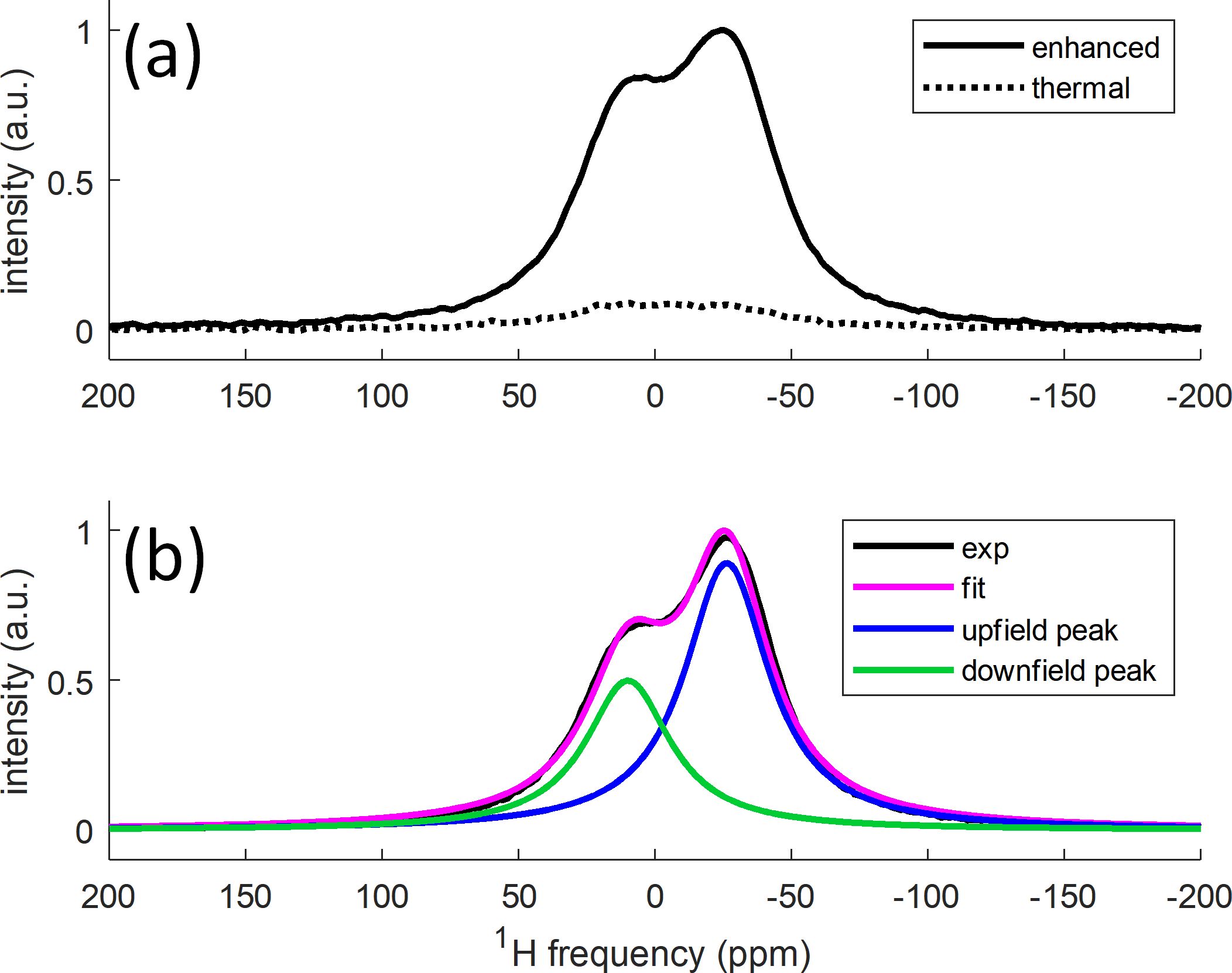}

\caption{a)$^{1}$H spectrum with DNP (MW irradiation at maximum of DNP spectrum) (solid) and without DNP (dashed), for Sample A1. b) The enhanced spectrum (black) fitted into two spectral lines (magenta); the upfield peak (blue) and the downfield peak (green).\label{fig:spectra}}
\end{center}
\end{figure}

\noindent Figure \ref{fig:spectra}a, shows the thermal (no microwave irradiation) and hyperpolarized $^{1}$H spectrum, obtained with microwave (MW) irradiation at the maximum of the DNP spectrum for 120 s, on Sample A1. At this MW frequency, the enhancement of the $^{1}$H nuclei is about 15, though it depends on the MW frequency, the MW irradiation time, the sample, and which NMR peak is considered. In this work, we define the enhancement as the ratio of the integrated intensity of each resonance, with and without MW irradiation.

Two peaks are observed in the spectrum, an ``upfield peak'' at -27 ppm and a ``downfield peak'' at 10 ppm. (Upfield means lower ppm values and downfield means higher ppm values). The referencing of the spectrum was performed in a separate experiment where we measured the Si powder and a D$_2$O sample (containing residual H$_2$O) at the same time. This experiment was used to assign the downfield peak to be at 10 ppm. Figure \ref{fig:spectra}b shows that the two lines can be fit using Lorenztian functions, with identical full widths at half of the maximum peak intensity (FWHM) of 5.4 kHz. The frequency of the downfield peak falls within the standard chemical shift range for $^{1}$H nuclei (and fits with -OH protons), whereas the upfield peak is highly shifted to negative ppm values. 

In principle, this large shift may be a paramagnetic shifts (such as Fermi-contact, pseudocontact or Knight Shift) due to interactions of the defect unpaired electrons with the $^{1}$H nuclei that make up the upfield peak, or due to bulk magnetic susceptibility (BMS) effects. \cite{RN340,RN341,RN342,RN349} 
Paramagnetic shifts are expected to be temperature dependent due to the interaction with the average electron polarization. To test for paramagnetic contributions to the shift, we measured NMR spectra from the sample as a function of temperature in the range between 4 K and 40 K, and observed that the separation between the peaks did not change with temperature in this range, ruling out such a paramagnetic contribution (see Supporting Information). Paramagnetic and BMS shifts are also orientation dependent, and would cause severe line broadening in powdered samples, which was not observed.

The observed shift may also be due to increased shielding of the $^{1}$H nuclei contributing to the upfield peak. Metal hydride systems (M-H, where M is a metal), have been known to have $^{1}$H chemical shifts in the range of 0 to -25 ppm, or even lower, due to large spin-orbit couplings or ring-currents within the d-orbitals of the heavy metal and transition metal atoms. \cite{RN351} Negative $^{1}$H chemical shifts are also expected in organic molecules where ring currents cause strong shielding of the $^{1}$H nuclei.\cite{RN352} While we do not understand the origin of the shifted peak, we speculate that it results from impurities at the surface that produce such a shielded environment.

Note that in previous experiments on the same powder only a single resonance was observed, with a FWHM of 6.2 kHz.\cite{RN163} In that work, however, after wetting the same with 20:80 H$_2$O:D$_2$O, the NMR line it had an additional shoulder indicating the appearance of a second upfield resonance. The separation between the two resonances was found to be 34 ppm, similar to the separation we see in this work.

\subsection{Intensity Variations}

\begin{figure}[H]
\begin{center}
\includegraphics[width=0.75\textwidth]{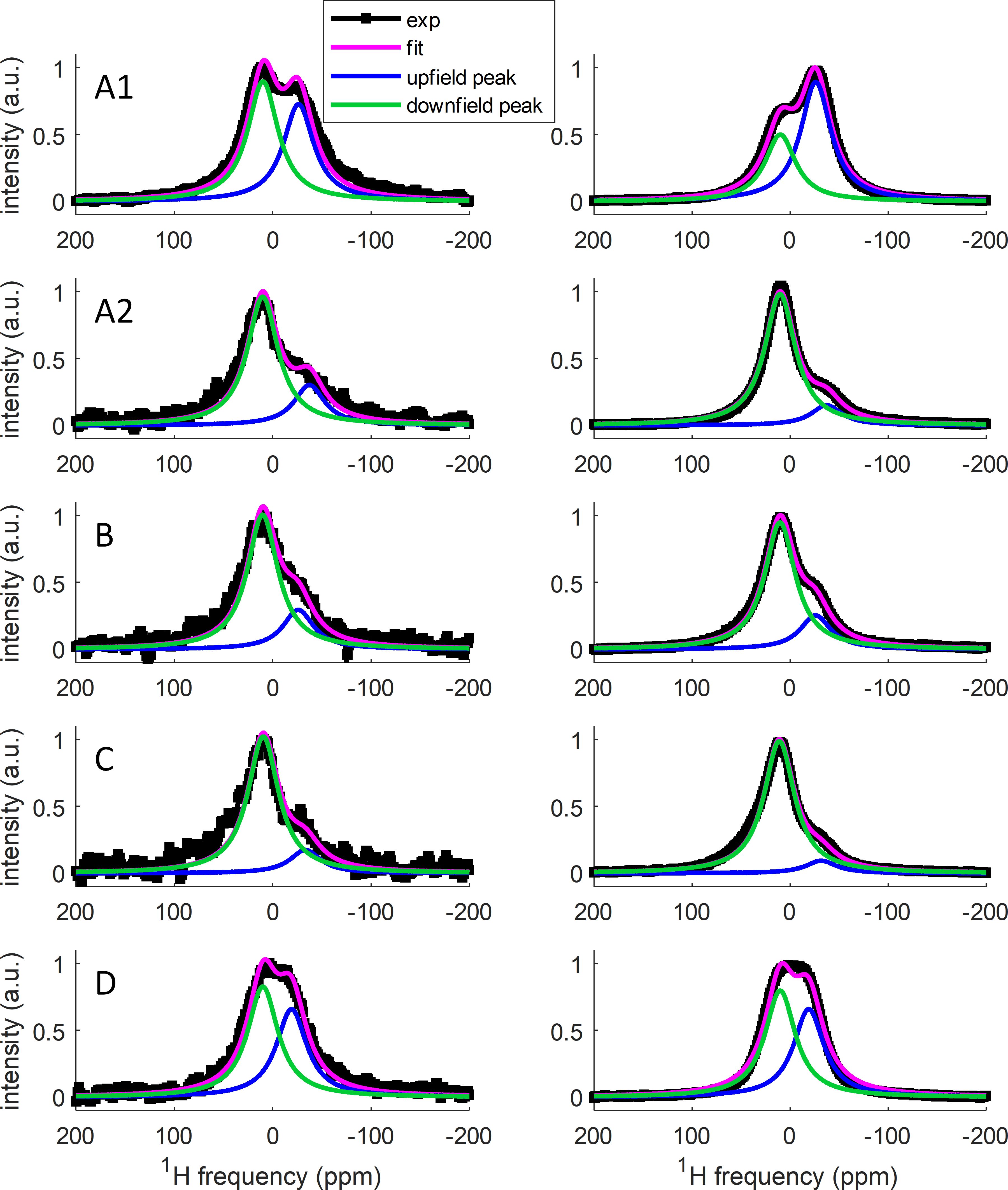}
\caption{Thermal (left column) and enhanced (right column) $^{1}$H spectra. In each row we plot the spectra for a different sample labeled A1, A2, B, C and D. Each panel contains an experimental spectrum (black squares), fitted using a deconvolution into two Lorentzian lines, with the total fit (magenta), the upfield peak (blue) and the downfield peak (green). The enhanced $^{1}$H spectra are plotted at the maximum of the DNP spectrum.
\label{fig:spectraNMRDNP}}
\end{center}
\end{figure}

\noindent In figure \ref{fig:spectraNMRDNP} we plot the thermal (MW-off) and hyperpolarized (MW-on) $^{1}$H spectra for all the samples measured. The $^{1}$H spectra were again fit to two Lorentzian lines as described above. More details on the fitting can be found in the Materials and Methods Section. The intensity of the downfield peak was observed to be fairly consistent across all measurements, whereas the intensity of the upfield peak varied across different experiments. The linewidth of both peaks was found to remain the same across all the experiments. The reason for the change in intensity is not clear, but suggests that the protons making up the upfield peak may be less tightly bound to the surface, resulting in a varying number of protons measured at the upfield site. The variations could also result from differences in the rate at which the system is cooled down from room temperature to 4 K.

In addition to changes in intensity, the separation between the two peaks was observed to vary between 29 ppm and 47 ppm in different experiments, potentially due to changes to the shielding experienced.
Small variations were observed in the absolute frequencies of the two peaks in different experiments - likely due to slight variations in the sample positioning in the coil. The homogeneity of the magnet used in the DNP experiment is on the order of 1 ppm as there are no room temperature shims in the bore. All shifts/assignments have thus been rounded to the nearest integer ppm values. As mentioned above the referencing of the spectra was performed in a different experiment. Since the downfield peak was observed to be relatively stable, we have assumed that this peak is at 10 ppm in all the experiments. The separation between the peaks was then used to assign the shift for the upfield peak. 

\subsection{Linewidth}
\noindent The 5.4 kHz FWHM of each resonance is very narrow for $^{1}$H nuclei under solid static conditions. We investigated the cause of this linewidth by measuring the T$_{2}$ of the lines in Sample A1 using both solid-echo and CPMG decay experiment. The solid echo refocuses dipolar interactions to lowest order as well as static magnetic field inhomogeneities, while the CPMG refocuses only the static field inhomogeneities. The CPMG sequence used an echo time of 200 \textmu s and pulse lengths of \textpi /2=0.5 \textmu s and \textpi =1 \textmu s.  

\begin{figure}[H]
\begin{center}
\includegraphics[width=0.7\textwidth]{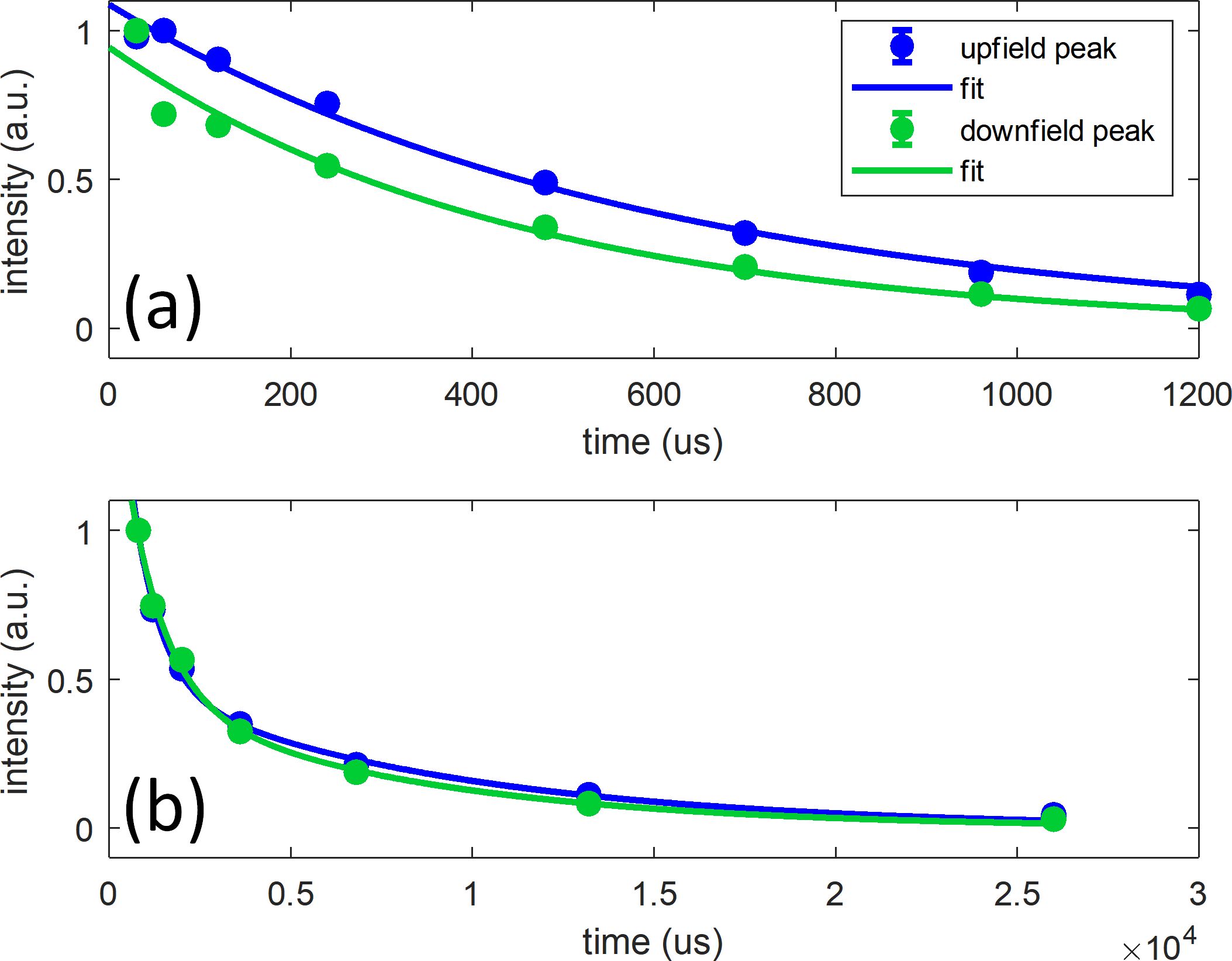}

\caption{Normalized T$_{2}$ curves measured by (top, symbols) varying the echo delay time in a solid echo pulse sequence, and by (bottom, symbols)
measuring CPMG with solid echo pulses, for the upfield peak (blue) and the downfield peak
(green): integrated intensity of each peak, plotted as a function
of the decay time. The exponential fits are given by the solid lines.
The error bars shown are given from the 95\% confidence interval of the integrated intensity of each peak given from fitting of the NMR spectrum for each point, and are smaller than the symbols in most cases. The MW irradiation was set at 93.8738 GHz, with 120 s irradiation for improved SNR.\label{fig:T2A1}}
\end{center}
\end{figure}

The solid echo decay was fit using a single exponential and the CPMG decay was fit using a double-exponential, with the short timescale about three times more dominant than the long timescale (see Figure \ref{fig:T2A1}). The long CPMG timescale is likely what is known as a \textquotedblleft CPMG tail\textquotedblright and is related to pulse imperfections\cite{RN277} and thus we will not consider it further. The decay times are given in Table \ref{tab:tableT2}.

\begin{table}[H]
\caption{The T$_2$ timescales measured using the solid echo and CPMG sequences,
with the amplitude of each CPMG timescale in parenthesis.\label{tab:tableT2}}

\begin{center}
\begin{tabular*}{0.65\textwidth}{@{\extracolsep{\fill}}|>{\centering}p{3cm}|>{\centering}p{3cm}|>{\centering}p{3cm}|}
\hline 
 & {\scriptsize{}T$_{2}$ (solid echo) (ms)} & {\scriptsize{}T$_{2}$ (CPMG) (ms)} \tabularnewline
\hline 
{\scriptsize{}A1, upfield peak} & {\scriptsize{}0.6\textpm 0.1} & {\scriptsize{}short: 0.7\textpm 0.5}{\scriptsize\par}

{\scriptsize{}(75\%\textpm 35\%)}{\scriptsize\par}

{\scriptsize{}long: 8.6\textpm 5.3}{\scriptsize\par}

{\scriptsize{}(25\%\textpm 15\%)} \tabularnewline 
\hline 
{\scriptsize{}A1,}{\scriptsize\par}

{\scriptsize{}downlfield peak} & {\scriptsize{}0.4\textpm 0.2} & {\scriptsize{}short: 1.0\textpm 0.9}{\scriptsize\par}

{\scriptsize{}(71\%\textpm 29\%)}{\scriptsize\par}

{\scriptsize{}long: 7.5\textpm 7.7 }{\scriptsize\par}

{\scriptsize{}(29\%\textpm 29\%)} \tabularnewline
\hline 
\end{tabular*}
\end{center}
\end{table}

The intrinsic linewidth for a Lorentzian peak corresponding to a decay time T$_2$ is $1/(2\pi T_2)$, so a T$_{2}$ of 0.5-1 ms corresponds to a width of 150-300 Hz, an order of magnitude lower than the observed linewidths. This suggests that the observed linewidths are dominated by inhomogeneous broadening. A dipolar coupling of 150 Hz correspond to a $^{1}$H-$^{1}$H distance of 70 nm. Alternatively, the homogeneous linewidth (determined by T$_2$) could be due to the electron T$_1e$, if the nuclear spins are isolated. The electron spin T$_{1e}$ for dangling-bond defects in silicon has been measured to be 0.04-40 ms depending on the sample oxidation at X-Band, which overlaps with the timescale of the observed nuclear spin T$_2$ in this work.\cite{RN350} Note that T$_{2}$ of Sample C was also measured and proved to the very similar to Sample A1 (data not shown).

\subsection{DNP Spectrum}

\begin{figure}[H]
\begin{center}
\includegraphics[width=0.65\textwidth]{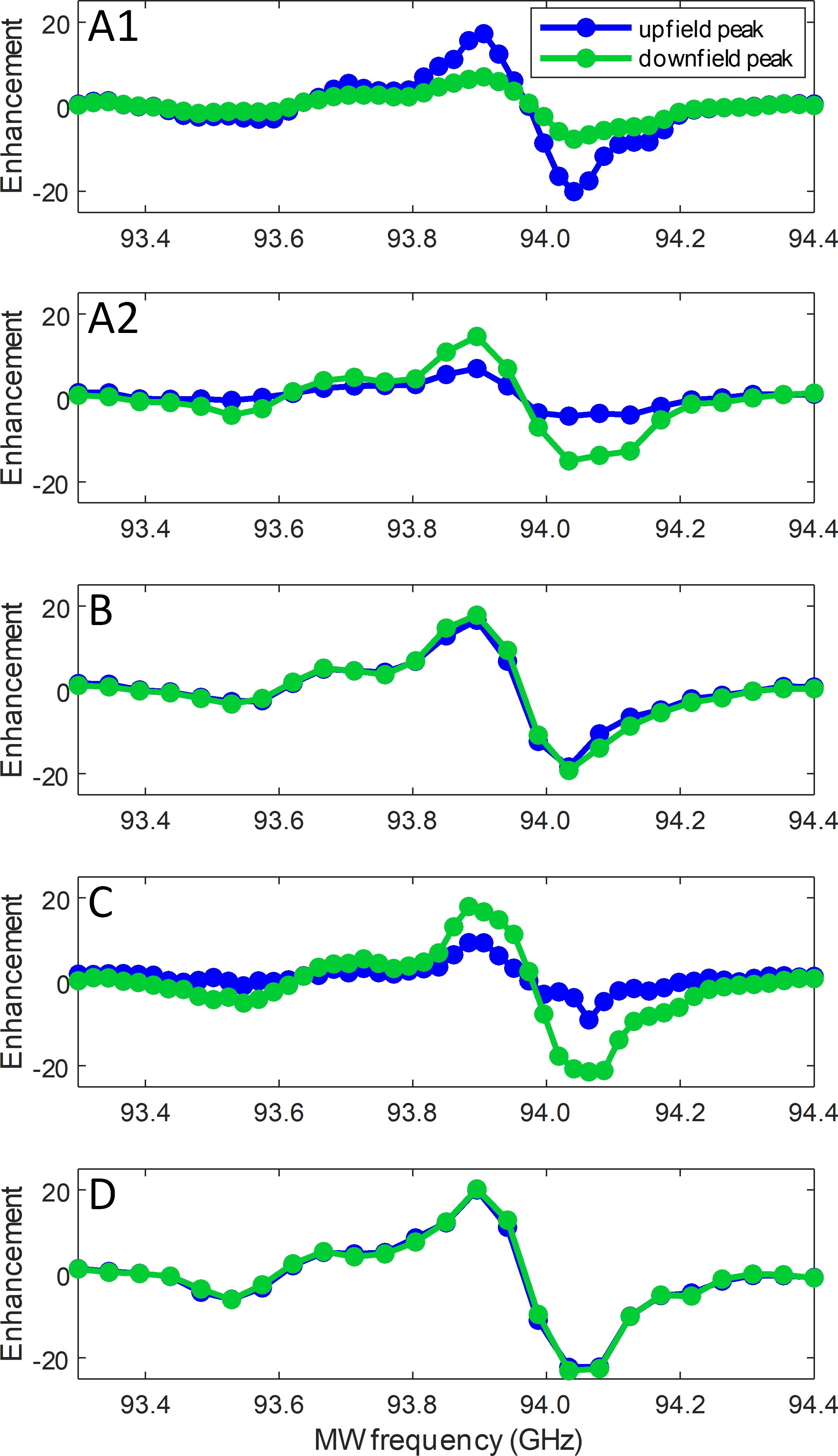}

\caption{DNP spectra for each sample used in this work (labeled A1, A2, B, C and D) plotted for the upfield peak (blue) and the downfield (green). \label{fig:DNPsweeps}}
\end{center}
\end{figure}

\noindent We studied the enhancement of both $^{1}$H peaks as a function of MW frequency (i.e. the DNP spectrum) as shown in Figure \ref{fig:DNPsweeps}.  The DNP spectrum was measured by setting the MW frequency and irradiation
length and varying the MW frequency with each subsequent experiment. The irradiation length was set to 120 s, as a compromise between larger enhancements and longer experimental times.

The DNP spectra for the upfield peak and the downfield peak have the same shape, but their relative enhancements vary for the different samples. For example, in Sample A1 the upfield peak is more enhanced than the downfield peak, but in Sample A2 it is the other way around. The enhancement of the downfield peak was observed to vary between samples much less than the enhancement of the upfield peak.  However, we do not see an obvious correlation between the relative intensities or separation between the upfield peak and the downfield peaks and the DNP enhancements.

The shape of the DNP spectrum is determined by both the DNP mechanism as well as the electron paramagnetic resonance (EPR) line of the electrons driving the enhancement.\cite{RN288,RN289} 
The solid effect (SE) enhancement appears when irradiating electron-nuclear  double quantum and zero quantum transitions appearing at $\omega_e \pm \omega_n$. This means that the SE DNP shape is expected to be broader than the EPR line, because it extends out past the edge of the line by $\omega_n$ on either side. \cite{RN305} The cross effect (CE), on the other hand, relies on pairs of electrons separated by $\omega_n$ frequency such that: $\omega_{e1} - \omega_{e2} = \omega_n$. \cite{RN289} \cite{RN306}  If these two electrons have different polarizations due to MW irradiation, they will enhance the nuclei. \cite{RN289} In this case, the enhancement from these electrons mainly appears within the EPR line, but can also extend out past the edge of the line by $\omega_n$ on either side in the case of the indirect CE (due to the effect of electron spectra diffusion). \cite{RN289}
The identical DNP shapes here likely means that electrons with similar g-anisotropy and EPR lines are the source of DNP enhancement for both types of $^{1}$H nuclei, and that the DNP mechanism is the same for both $^{1}$H nuclei. 

\begin{figure}[H]
\begin{center}
\includegraphics[width=0.65\textwidth]{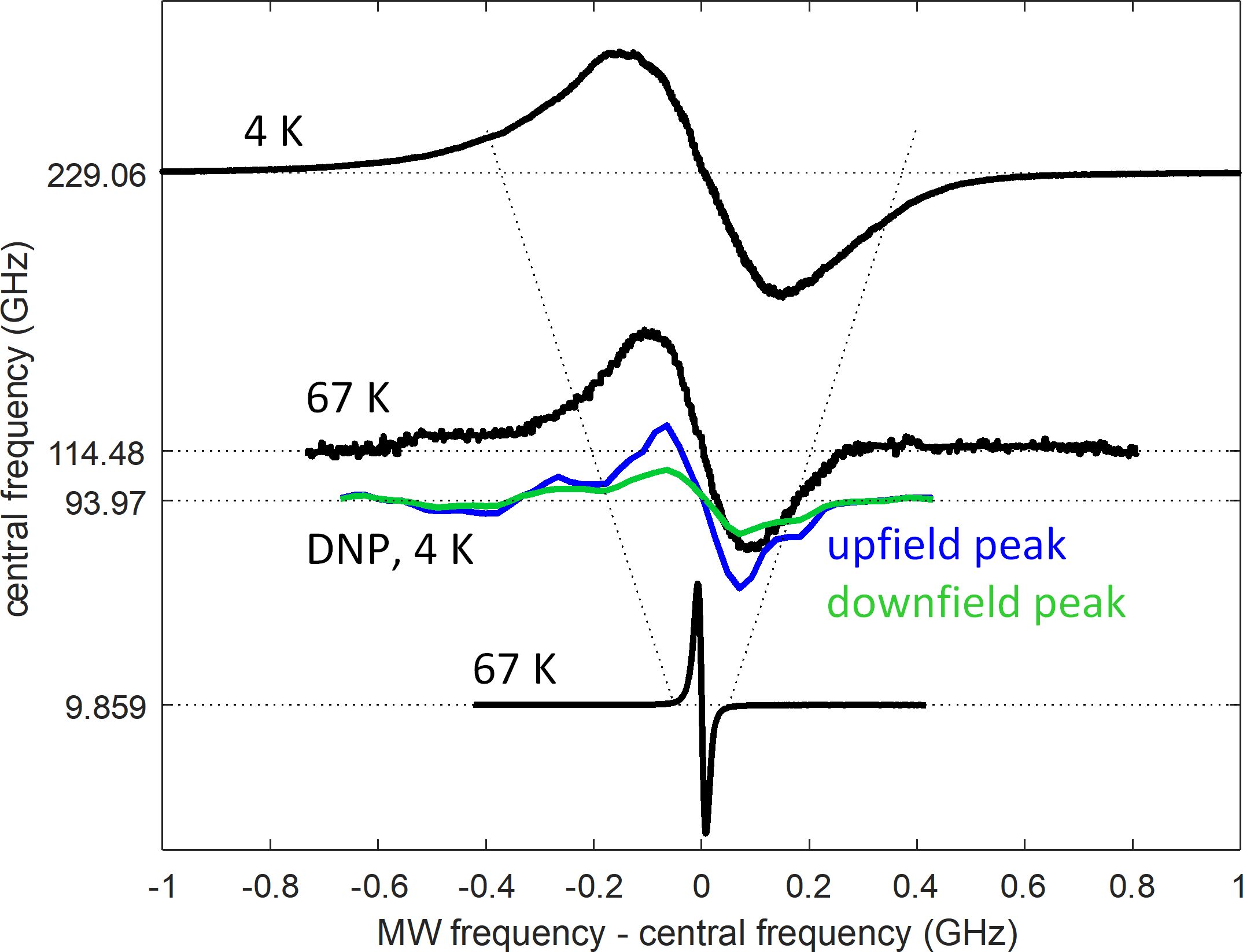}

\caption{EPR spectra of the silicon particles as a function of field. Each spectrum is referenced to the central frequency at 0 GHz. The y-axis represents the central frequency. Plotted are spectra at 229.06 GHz at 4K, 114.48 GHz at 67 K and 9.859 GHz at 67 K. Also plotted are the DNP spectra of Sample A1, measured at 93.97 GHz and 4 K. The horizontal dashed lines mark the baseline of each spectrum, and the diagonal dashed lines show the width of the EPR line as it changes linearly with the field. \label{fig:EPR}}
\end{center}
\end{figure}

The EPR line of the silicon particles was measured at 9 GHz, 115 GHz and 230 GHz, and can be seen in Figure \ref{fig:EPR}. The X-Band spectrum was measured on a commercial Bruker EMX spectrometer. The high field spectra were measured by a high frequency EPR spectrometer at University of Southern California.\cite{RN354} In the experiments, the HF microwave power and the magnetic field modulation strengths were adjusted carefully to perform EPR measurements in the linear regime where the saturation effect is negligible.\cite{RN355} 

Figure~\ref{fig:EPR} shows the EPR spectra observed at the three different frequencies, as well as the DNP spectrum for sample A1.  Only a single EPR line with g$\sim$2 is observed in all cases. The width of the EPR line is clearly seen to scale with field (as shown by the dotted lines), indicating that it is dominated by inhomogeneous broadening. The EPR spectra shown were measured at either 4 K or 67 K. No variation in the EPR lineshape was observed with temperature (see Supporting Information). Note also that these spectra are similar to the EPR lineshapes published by Cassidy et al. and Guy et al., measured at X-Band and W-Band, respectively, for the same particles. \cite{RN335,RN163}

The observed width of the EPR line is approximately 600 MHz at 115 GHz, suggesting an inhomogeneous linewidth of 490 MHz at 94 GHz.  Figure \ref{fig:EPR} show that most of the observed DNP enhancement falls within the width of the EPR line (between -170 MHz and 140 MHz), but that there is still enhancement outside (<-170 GHz and >140 GHz).
Considering this, we assign the DNP enhancement features to a combination of the CE-DNP mechanism and the SE-DNP mechanism. \cite{RN288,RN289} The CE mechanism (direct or indirect) \cite{RN289} is possible because silicon particles are known to have a typical surface electron concentration of 10$^{12}$ cm$^{-2}$ or higher,\cite{RN336,RN337,RN338} which corresponds to electron-electron dipolar interactions of 50 KHz on average. There are also small features around -420 MHz in Figure \ref{fig:EPR} which may be due to higher order DNP effects. \cite{RN326}

It should be noted that the above DNP spectra were recorded using a 120 s MW irradiation, which is shorter than the characteristic build-up time for the DNP experiment. We measured both the spin-lattice relation time T$_{1n}$ and the build up time T$_{bu}$ for both peaks for samples A1 and C.  The nuclear spin-lattice relaxation time, T$_{1n}$, was measured using a saturation recovery experiment, beginning with a train of saturation pulses, and then a delay which varied to allow the nuclear polarization to recover. The DNP buildup, T$_{bu}$, was measured in the exact same manner, but with MW irradiation applied during the recovery delay period.

The data were fit to a stretched exponential function, which represents a distribution of relaxation times or polarization buildup times (see Supporting Information). A stretching coefficient ($\beta = 0.8$) was used in all cases. All the the characteristic build up times lie in the range of 300-400 s, as can be seen in Table \ref{tab:table-2}. Thus the enhancements at 120 s MW irradiation are likely to reflect the steady-state enhancements at longer times. There is greater uncertainty in the values of T$_{1n}$ due to the low signal to noise ratio in the thermal experiments. We observed that T$_{bu}$<T$_{1n}$ in all cases, as expected.

\begin{table}[H]
\caption{The stretched exponential timescales of T$_{1n}$ and T$_{bu}$ (\textbeta=0.8) and the maximum DNP enhancement observed at 120 s. \label{tab:table-2}}

\begin{center}
\begin{tabular*}{0.8\textwidth}{@{\extracolsep{\fill}}|>{\centering}p{2.5cm}|>{\centering}p{2.5cm}|>{\centering}p{2.5cm}|>{\centering}p{2.5cm}|}
\hline 
 &  {\scriptsize{}T$_{1n}$ (s)} & {\scriptsize{}T$_{bu}$(s)} & {\scriptsize{}Enhancement}\tabularnewline
\hline 
{\scriptsize{}A1, upfield peak}  & {\scriptsize{}486$\pm$95} & {\scriptsize{}369.7$\pm$43.5} & {\scriptsize{}15.6}\tabularnewline
\hline 
{\scriptsize{}A1,}{\scriptsize\par}

{\scriptsize{}downlfield peak} & {\scriptsize{}913.4$\pm$405.9}{\scriptsize\par}

 & {\scriptsize{}305.7$\pm$37.2} & {\scriptsize{}6.5}\tabularnewline
\hline 
{\scriptsize{}C, }{\scriptsize\par}

{\scriptsize{}upfield peak} & {\scriptsize{}1831$\pm$1618} & {\scriptsize{}305.2$\pm$105.1}{\scriptsize\par}

 & {\scriptsize{}7.0}\tabularnewline
\hline 
{\scriptsize{}C, }{\scriptsize\par}

{\scriptsize{}downfield peak} &  {\scriptsize{}638.6$\pm$201.7} & {\scriptsize{}401.7$\pm$27.3} & {\scriptsize{}14.6}\tabularnewline
\hline 
\end{tabular*}
\end{center}
\end{table}

We also explored the role of frequency modulation of the MW irradiation on Sample B, where the upfield peak was much smaller than the downfield peak.  The frequency of the modulation was 10 kHz, and the amplitude was 70 MHz (\textpm 70 MHz around the central frequency) in all cases.
The frequency of the modulation refers to how many times per second the frequency range was swept. The amplitude  of the modulation refers to the range of MW frequencies that was swept around the central frequency. Three types of frequency modulation were used: sawtooth up, sawtooth down and symmetrical triangular modulation. For DNP spectra the enhancement is plotted at each central frequency.

\begin{figure}[H]
\begin{center}
\includegraphics[width=0.7\textwidth]{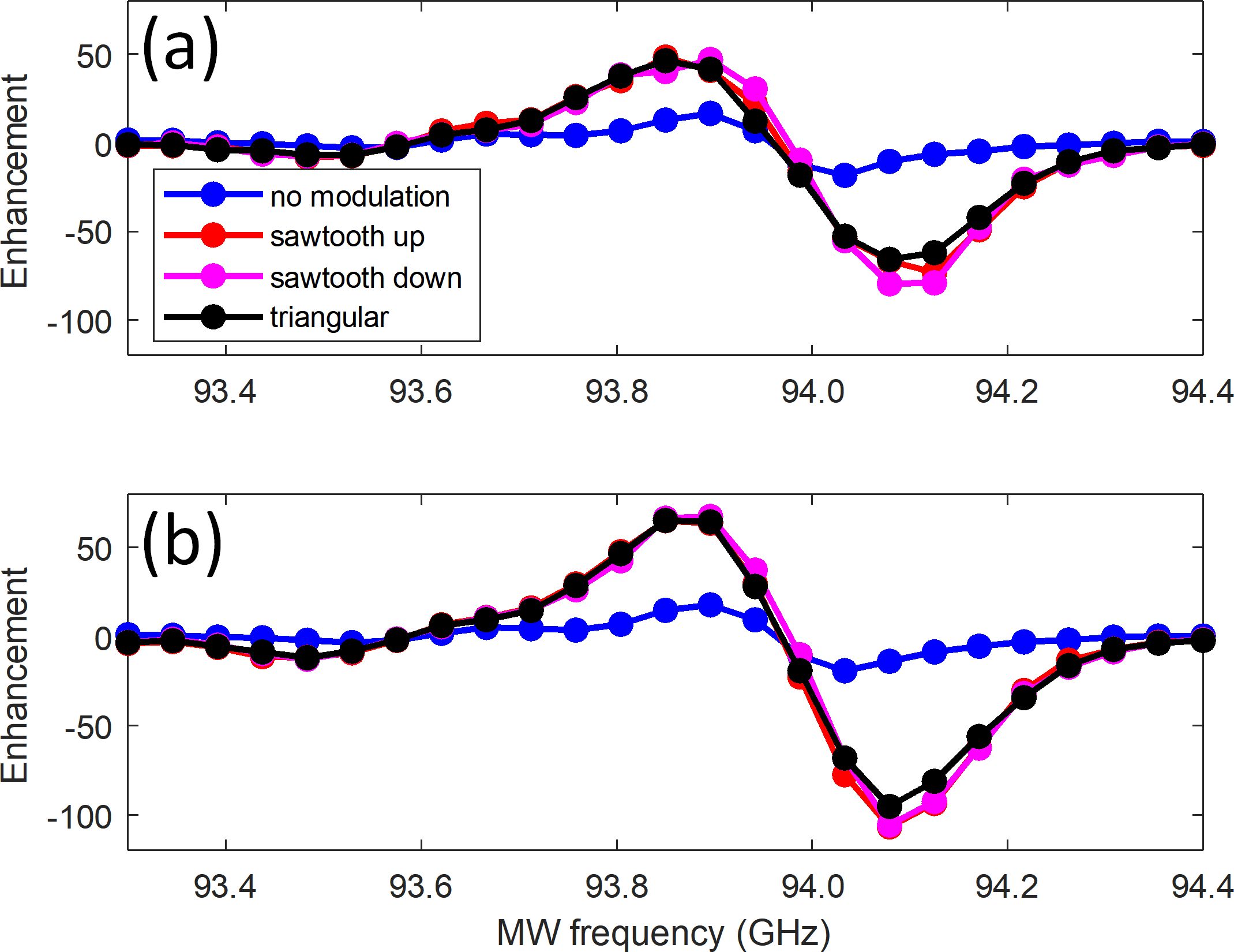}
\caption{DNP spectrum of (a) the upfield peak and (b) the downfield peak of sample B: integrated
intensity of each peak divided by the integrated intensity of the
thermal MW off signal, plotted as a function of the MW irradiation
frequency, with 120 s of irradiation at each frequency. Different
modulation schemes are compared: no modulation (blue), sawtooth up
(red), sawtooth down (magenta) and triagonal/symmetric (black). In
all cases, the modulation frequency was 10 kHz and the modulation
amplitude was $\pm70$ MHz around the center frequency.\label{fig:modulation}}
\end{center}
\end{figure}

The frequency modulation results are plotted in Figure \ref{fig:modulation}, with the upfield peak in panel (a) and the downfield peak in panel (b). All forms of modulation resulted in significantly increased enhancement of both the upfield and the downfield peaks, compared to no modulation, as previously shown. \cite{RN163} \cite{RN197} \cite{RN249} 

The upfield peak showed a lower enhancement in the negative part of the DNP spectrum (~94.1 GHz) when triangular modulation was applied, relative to the single-directional sawtooth sweeps. This may be because the frequency of the triangular modulation is effectively twice that of sawtooth modulation (see Supporting Information), because it has been shown but that when the modulation frequency is much faster than the electron spin-lattice relaxation time T$_{1e}$ a decrease in enhancement can be expected. \cite{RN163} However, it is a small effect, and not clear why the upfield peak is affected more than the downfield peak.

\subsection{High-field DNP}
\noindent To learn more about the proton spectrum
we measured the $^{1}$H NMR spectrum on a 600 MHz Bruker Avance III DNP spectrometer, with 8 kHz magic angle spinning (MAS) at a temperature of 100K. Experiments with MW irradiation were conducted with a collector current of 110 mA, with irradiation at 395.287 GHz. $^{1}$H presaturation was done for all experiments with a train of 200 pulses of 2.6  \textmu s  with a 1 ms delay between each pulse.

The $^{1}$H signal was collected with a 16 step background  suppression, with a \textpi /2=2.6 \textmu s excitation and a composite pulse for background suppression. \cite{RN285} Sixteen scans were acquired, with a 60 s recycle delay between every transient.

\begin{figure}[H]
\begin{center}
\includegraphics[width=0.7\textwidth]{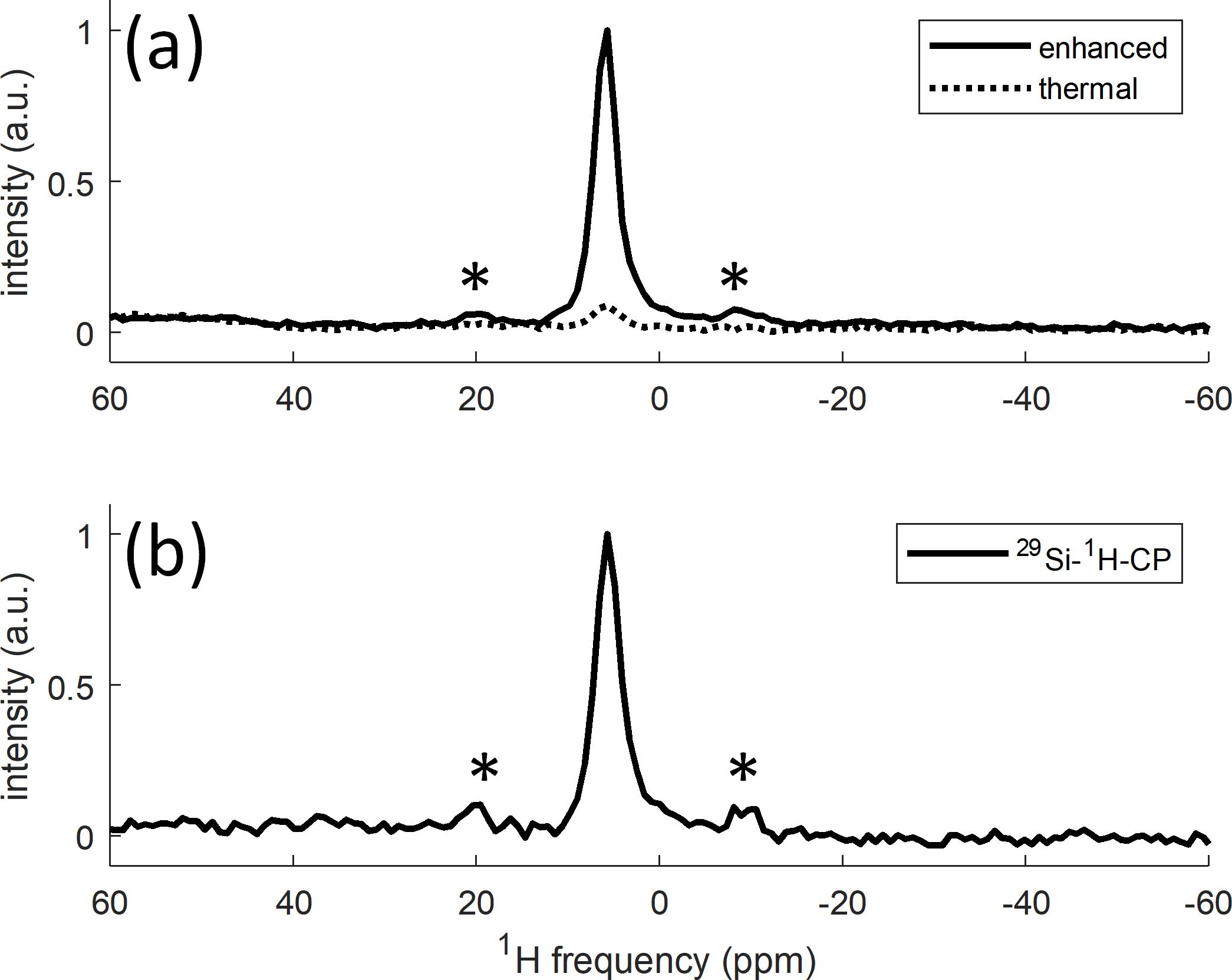}

\caption{a) $^{1}$H spectrum with DNP (solid)
and without DNP (dashed), with 8 kHz MAS, measured on a 600 MHz DNP
spectrometer. b)$^{29}$Si-$^{1}$H-CPMAS spectrum with DNP, with 8 kHz MAS, measured on a 600 MHz DNP spectrometer. The MW irradiation frequency was 395.287 GHz.
Spinning sidebands are marked with {*}.\label{fig:spectra-nottingham}}
\end{center}
\end{figure}

At the higher field, only a single narrow resonance is detected at 5.7 ppm, with a FWHM of 1.7 kHz. The line measured at 600 MHz appears at 5.7 ppm, which is closer to the 10 ppm line observed at 3.3 T. The spectra were referenced to $^1$H adamantane. The linewidth of the peak measured with MAS at 600 MHz is close to 2 kHz, however, the spinning sidebands indicate that the static linewidth is likely closer to 5 kHz, which is a similar linewidth to that measured at 142 MHz. In Figure \ref{fig:spectra-nottingham}a we compare the DNP enhanced signal to the thermal signal, showing the enhancement of 11.3. 

In Figure \ref{fig:spectra-nottingham}b we show the
$^{1}$H spectrum detected via inverse cross polarization (CP) from
$^{29}$Si to $^{1}$H with MW irradiation (directly enhancing the $^{29}$Si). This spectrum is identical in shape to the spectrum with direct $^{1}$H detection, indicating that the $^{1}$H
nuclei in the narrow line, which are on the surface of the particles, are dipolar coupled with $^{29}$Si nuclei. The cross polarization experiments from $^{29}$Si to $^{1}$H with MAS ($^{29}$Si-$^{1}$H-CPMAS) \cite{RN273} were conducted with MW irradiation of 60 s such that the $^{29}$Si nuclei were directly enhanced. A contact time of 2 ms, and a spin-locking strength of 96 kHz were used for the CP. A \textpi /2=3 \textmu s pulsed was used for $^{29}$Si excitation and 1024 scans were acquired. 

Note, that a static spectrum
was attempted but only 16 scans were acquired, and this was not enough to detect signal. This could be because the T$_{1n}$ was longer under static conditions, and the delay between scans was not sufficient.

\section{Conclusions}

\noindent In this work, we show DNP enhancement of $^{1}$H nuclei on the surface of silicon microparticles using only endogenous defect unpaired electrons as polarization sources. Using static DNP, we are able to get enhancements of up to 55 without the need for additional radicals or solvents. We see two $^{1}$H resonances, separated by a large shifts ranging from 29 to 47 ppm whose origin is unclear. The two types of $^{1}$H nuclei have homogeneous linewidths on the order of 150-200 Hz, indicating they are relatively sparse. The intensity of the "upfield peak" is highly variable, while the intensity of the "the downfield peak" is much more stable. Despite very similar relaxation times (T$_{1n}$ and T$_{2}$) and linewidths, the two resonances are enhanced to different degrees. These results suggest that there are two distinct microscopic environments on the surface of the Si particles where the protons reside.

\section{Materials and Methods}

\noindent For more information about how the DNP spectrometer, as well as experimental details of the DNP spectra, T$_2$, T$_{1n}$ and T$_{bu}$ curves see the main text.

\subsection{Sample preparation:}
\noindent  All samples were prepared from the same batch of silicon particles from Alfa Aesar, and were used as is, as described in the main text. Samples A1, A2, B and D were cooled gradually to 4 K over the course of several hours. However, sample C was directly inserted into a cryostat that was already at 4 K.

\subsection{Variable temperature experiments:}

\noindent The $^{1}$H Si powder spectrum was measured as a function of temperature
from 40 K to 4.2 K, and the $^{1}$H spectrum of residual H$_{2}$O
in D$_{2}$O was measured from 50 K to 4.2 K. At each temperature
the $^{1}$H spectrum was collected via the same solid echo detection
described above.

\subsubsection{X-Band cw-EPR lines:}

\noindent X-Band continuous-wave (cw) electron paramagnetic resonance (EPR) line was measured on a Bruker EMX spectrometer. The MW frequency was set to 9.834 GHz, with power set to 0.128 mW. Frequency modulation of 100 kHz was used, with an amplitude of 1.5 G. A range of 200 G was swept, centered around 3520 G, with a resolution of 512 points.

\subsection{High field cw-EPR lines:}

\noindent Continuous-wave (cw) electron paramagnetic resonance (EPR) lines were measured at 115 GHz and 230 GHz.  
\begin{itemize}
    \item At 115 GHz:
The MW frequency was set to 115 GHz, with power set to 200 \textmu W. Frequency modulation of 20 kHz was used, with an amplitude of 0.02 mT. Temperatures between 67 K and 250 K were recorded.

\item At 230 GHz:
The MW frequency was set to 230 GHz, with power set to 50 \textmu W. Frequency modulation of 20 kHz was used, with an amplitude of 0.02 mT. Temperatures between 4 K and 20 K were recorded.
\end{itemize}
\subsection{Data processing:}

\noindent The $^{1}$H signal was fit to two Lorentzian lines. In all cases the peaks were fit with a linewidth of 5.4 kHz, which is the width that give the best fit for Sample A1. The frequencies of the peaks were determined for each sample separately using the highest signal to noise ratio (SNR) spectrum within the set of data (recorded in one day and during one cool-down). The frequencies extracted from this fit were then used to fit all other data in the set, using the pre-determined widths, only allowing the amplitude to vary. For the DNP spectra, a different phase correction was applied at each MW frequency, in order to ensure the correct phasing of the spectra.

For the variable temperature experiments, the spectra were fit either with two Lorentzian lineshapes or one Lorentzian lineshape,
for the Si powder or the H$_{2}$O, respectively. The linewidths were
kept constant and determined by the lineshape at 4.2 K, while the
frequency and amplitude of the lines were allowed to vary in order to fit the data as a function of temperature.

In some cases, when the NMR intensity is low, a broad $^{1}$H resonance
is observed, which we attribute to probe background signal. The amplitude
of the background is minimized by using an echo delay of \texttau =200
\textmu s, and can be neglected during line fitting. 

The spectra
were referenced to tetramethylsilane (TMS) at 0 ppm, using a secondary
reference of residual H$_{2}$O in D$_{2}$O at 4.9 ppm, also measured
at 4 K. This resulted in the ``downfield peak'' appearing at 10 ppm.

The T$_\mathrm{1n}$ and T$_\mathrm{bu}$ curves were fitted with a stretched-exponential function:

\begin{equation}
y=M_\mathrm{eq}\left(1-e^{-(t/T_\mathrm{1n})^{0.8}}\right) \label{eq:T1n}
\end{equation}
\begin{equation}
y=M_\mathrm{eq}\left(1-e^{-(t/T_\mathrm{bu})^{0.8}}\right) \label{eq:Tbu}
\end{equation}
where T$_\mathrm{1n}$ is the relaxation time, T$_\mathrm{bu}$ is the enhancement buildup time and $\beta=0.8$ is the scaling factor that determines the shape of the curve. The scaling factor was constrained to 0.8 to facilitate comparison between the timescales. This value was determined empirically by fitting all the curves and then choosing a value of $\beta$ that results in good fits for all curves.

The T$_2$ curves were fitted with a single-exponential or a double-exponential
function:

\begin{equation}
y=M_{0}^\mathrm{long}e^{-t/T_{2}^\mathrm{long}}+M_{0}^\mathrm{short}e^{-t/T_{2}^\mathrm{short}}\label{eq:T2}
\end{equation}
Where T$_{2}^\mathrm{long}$ and T$_{2}^\mathrm{short}$ are the long and short
timescale respectively, and $M_{0}^\mathrm{long}$ and $M_{0}^\mathrm{short}$ are
their amplitudes. Only a single term in Equation \ref{eq:T2}
was used for single-exponential fitting.

\subsection{Scanning Electron Microscope:}
\noindent The scanning electron microscope (SEM) image in the Supporting Information was taken
on a FEI (Thermo Fisher Scientific) Scios2 LoVac dual beam FEG/FIB SEM, using
a T1 detector, an acceleration voltage of 5kV, a current of 0.8 nA and a magnification of 2500 x.
The silicon particles were mounted to a specimen stage using double-sided tape.

\section*{Acknowledgements:}
\noindent This research was funded in part by the US National Science Foundation under Grant Nos. CHE-1410504 (CR), DMR-1508661 (S.T.) and CHE-1611134) (S.T.), the Nottingham University Engineering and Physical Sciences Partnered Access Fund and a Dartmouth College Global Exploratory and Development Grant. The Nottingham DNP MAS NMR Facility is funded by Grants EP/R042853/1 and  EP/L022524/1. We would like to thank Dr. Maxime J. Guinel for his help with the SEM experiments. 

\bibliography{references_Si_paper}

\end{document}


\begin{frontmatter}
\title{Supporting Information for \\ DNP-NMR of surface hydrogen on silicon particles}

\author{Daphna Shimon$^a$}
\ead{daphna.shimon@dartmouth.edu}
\author{Kipp J. van Schooten$^a$}
\address[1]{Department of Physics and Astronomy, Dartmouth College, Hanover,
NH 03755, U.S.A.}
\author{Subhradip Paul$^b$}
\address[2]{School of Physics and Astronomy, University of Nottingham, Nottingham NG7 2RD, UK.}
\author{Zaili Peng$^c$}
\author{Susumu Takahashi$^{c,d}$}
\address[3]{Department of Chemistry, University of Southern California, Los Angeles, California 90089, USA}
\address[4]{Department of Physics and Astronomy, University of Southern California, Los Angeles, California 90089, USA}
\author{Walter K\"ockenberger$^b$}
\author{Chandrasekhar Ramanathan$^a$}
\ead{chandrasekhar.ramanathan@dartmouth.edu}
\end{frontmatter}

\section{EPR as a function of temperature}

\noindent We have measured the EPR lines at 115 GHz and 230 GHz as a function of temperature, showing that there is only a slight broadening of the lines at lower temperatures.

\begin{figure}[H]
\begin{center}
\includegraphics[width=0.8\textwidth]{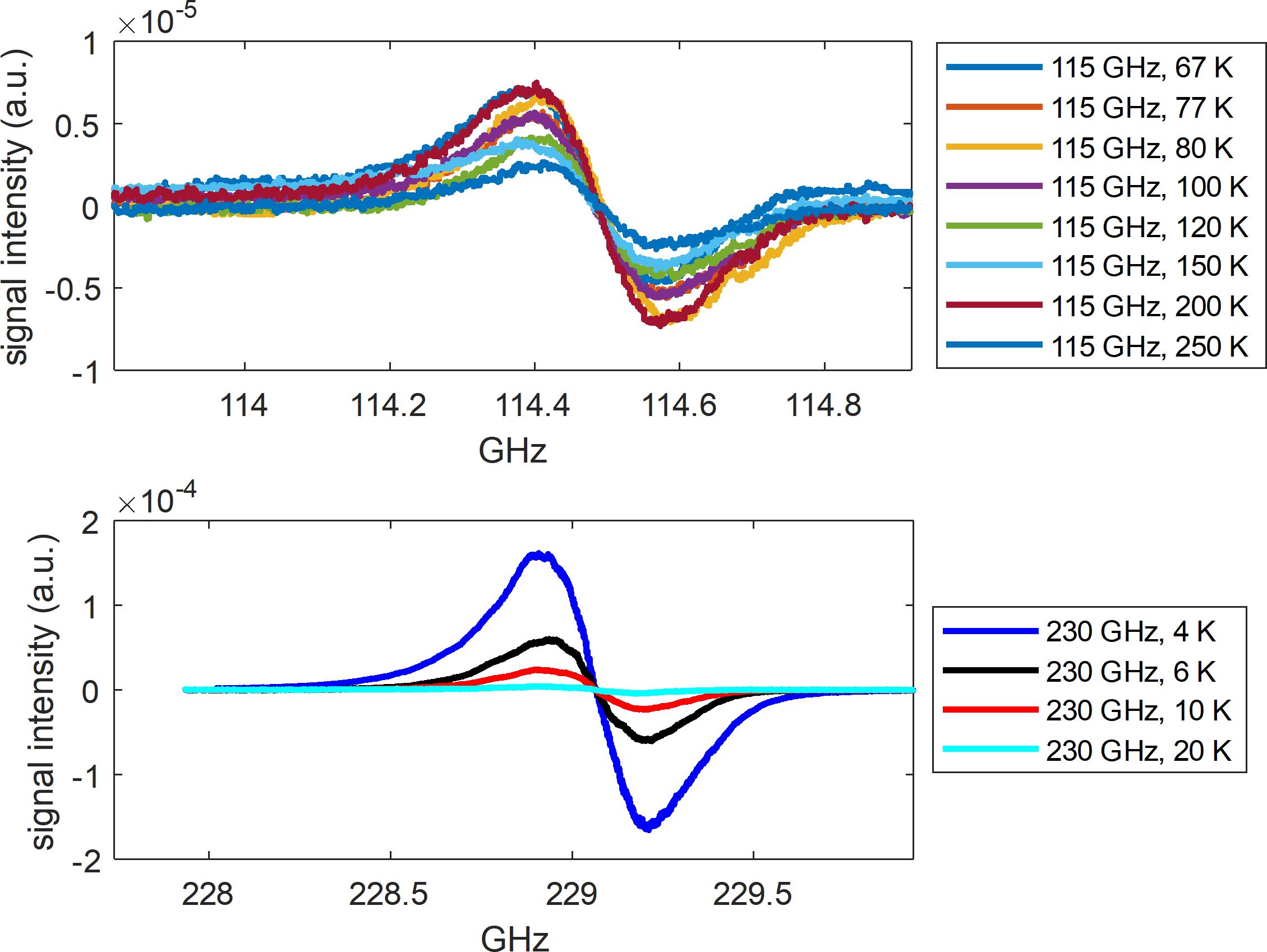}

\caption{EPR lines measured at 115 GHz and 230 GHz as a function of temperature. \label{EPR_vs_temp}}
\end{center}
\end{figure}

\section{T$_{1n}$ and T$_{bu}$ curves}

\noindent It is known that longer T$_{1n}$ can result in higher enhancement, and that T$_{bu}$ is limited by T$_{1n}$ and reflects the effective strength of the DNP transitions and the efficiency of the spin diffusion (i.e. the spread of enhancement throughout the sample).\cite{RN288} \cite{RN305} \cite{RN306} \cite{RN328}

\begin{figure}[H]
\begin{center}
\includegraphics[width=0.65\textwidth]{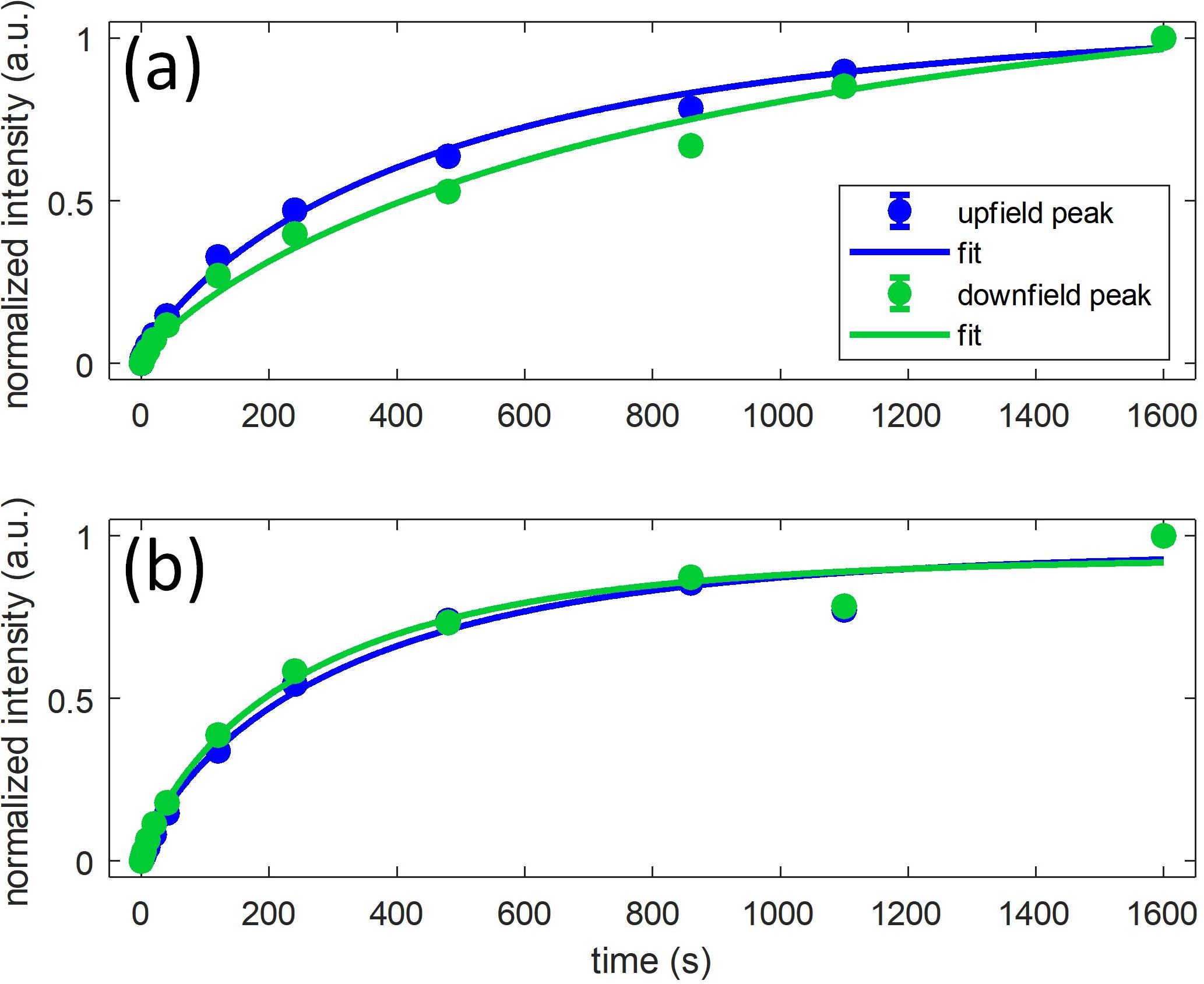}

\caption{Sample A1 saturation recovery T$_{1n}$ curves (top row, symbols) and MW buildup
T$_{bu}$ (bottom row, symbols) for the upfield peak (blue) and the downfield peak (green):
integrated intensity of each peak, plotted as a function of the recovery
time. The exponential fits of the normalized curves
are given by the solid lines. The error bars shown are given from
the 95\% confidence interval of the integrated intensity of each peak
given from fitting of the NMR spectrum for each point, and are smaller
than the symbols in most cases.\label{fig:T1nA1}}
\end{center}
\end{figure}

\begin{figure}[H]
\begin{center}
\includegraphics[width=0.65\textwidth]{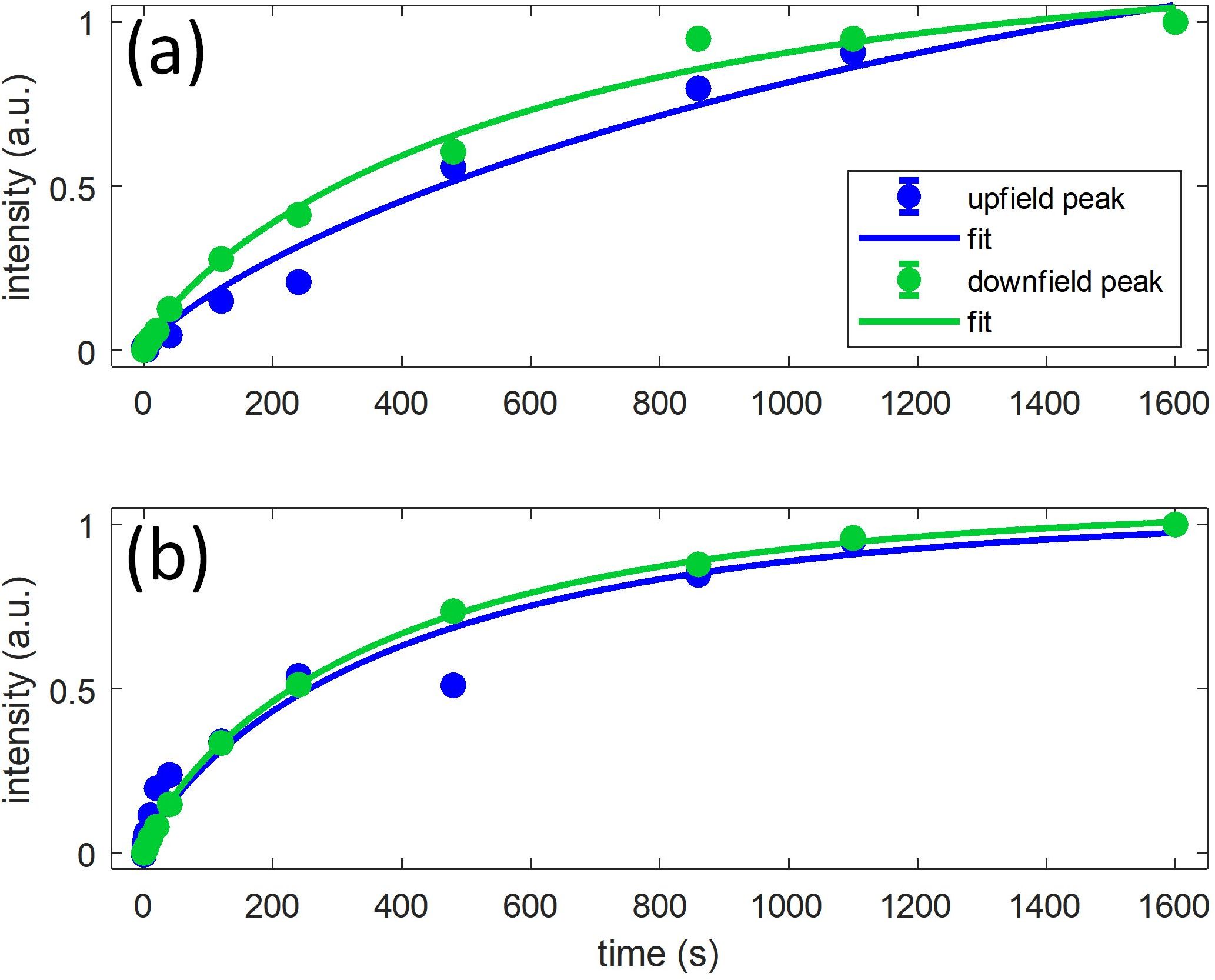}

\caption{Sample C saturation recovery T$_{1n}$ curves (top row, symbols) and MW buildup T$_{bu}$ (bottom row, symbols) for the upfield peak (blue) and the downfield peak (green): integrated intensity of each peak, plotted as a function of the recovery time. The exponential fits of the normalized curves are given by the solid lines. The error bars shown are given from
the 95\% confidence interval of the integrated intensity of each peak
given from fitting of the NMR spectrum for each point, and are smaller
than the symbols in most cases.\label{fig:T1nC}}
\end{center}
\end{figure}

We measured the spin-lattice relaxation time T$_{1n}$ and the DNP buildup time, T$_{bu}$, for samples A1 and C.  The data, along with the stretched-exponential fits are shown in Figure \ref{fig:T1nA1} and \ref{fig:T1nC}. We used a stretching parameter $\beta = 0.8$ for all the fits.

The fit parameters are summarized in Table 2 in the main paper and below.  All the characteristic build up times lie in the range of 300-400 s, as can be seen in Table~\ref{tab:table-2}. Given this similarity in the T$_{bu}$ values, the variation in DNP enhancement observed at the fixed growth time of 120 s is likely to reflect the variation in the steady state enhancement.

There is greater uncertainty in the values of T$_{1n}$ due to the low signal to noise ratio in the thermal experiments. We observed that T$_{bu}$<T$_{1n}$ in all cases, as expected.  None of the curves fully reach the steady state enhancement because the experiments were prohibitively long.

\begin{table}[H]
\caption{The stretched exponential timescales of T$_{1n}$ and T$_{bu}$ (\textbeta=0.8) and the maximum DNP enhancement observed at 120 s. \label{tab:table-2}}

\begin{center}
\begin{tabular*}{0.65\textwidth}{@{\extracolsep{\fill}}|>{\centering}p{3cm}|>{\centering}p{3cm}|>{\centering}p{3cm}|}
\hline 
 &  {\scriptsize{}T$_{1n}$ (s)} & {\scriptsize{}T$_{bu}$(s)} \tabularnewline
\hline 
{\scriptsize{}A1, upfield peak}  & {\scriptsize{}486$\pm$95} & {\scriptsize{}369.7$\pm$43.5} \tabularnewline
\hline 
{\scriptsize{}A1, downfield peak} & {\scriptsize{}913.4$\pm$405.9}{\scriptsize\par}

 & {\scriptsize{}305.7$\pm$37.2} \tabularnewline
\hline 
{\scriptsize{}C, upfield peak }& {\scriptsize{}1831$\pm$1618} & {\scriptsize{}305.2$\pm$105.1}{\scriptsize\par}

\tabularnewline
\hline 
{\scriptsize{}C, downfield peak} &  {\scriptsize{}638.6$\pm$201.7} & {\scriptsize{}401.7$\pm$27.3} \tabularnewline
\hline 
\end{tabular*}
\end{center}
\end{table}




\section{Temperature dependence of NMR lines}

\begin{figure}[H]
\begin{center}
\includegraphics[width=0.7\textwidth]{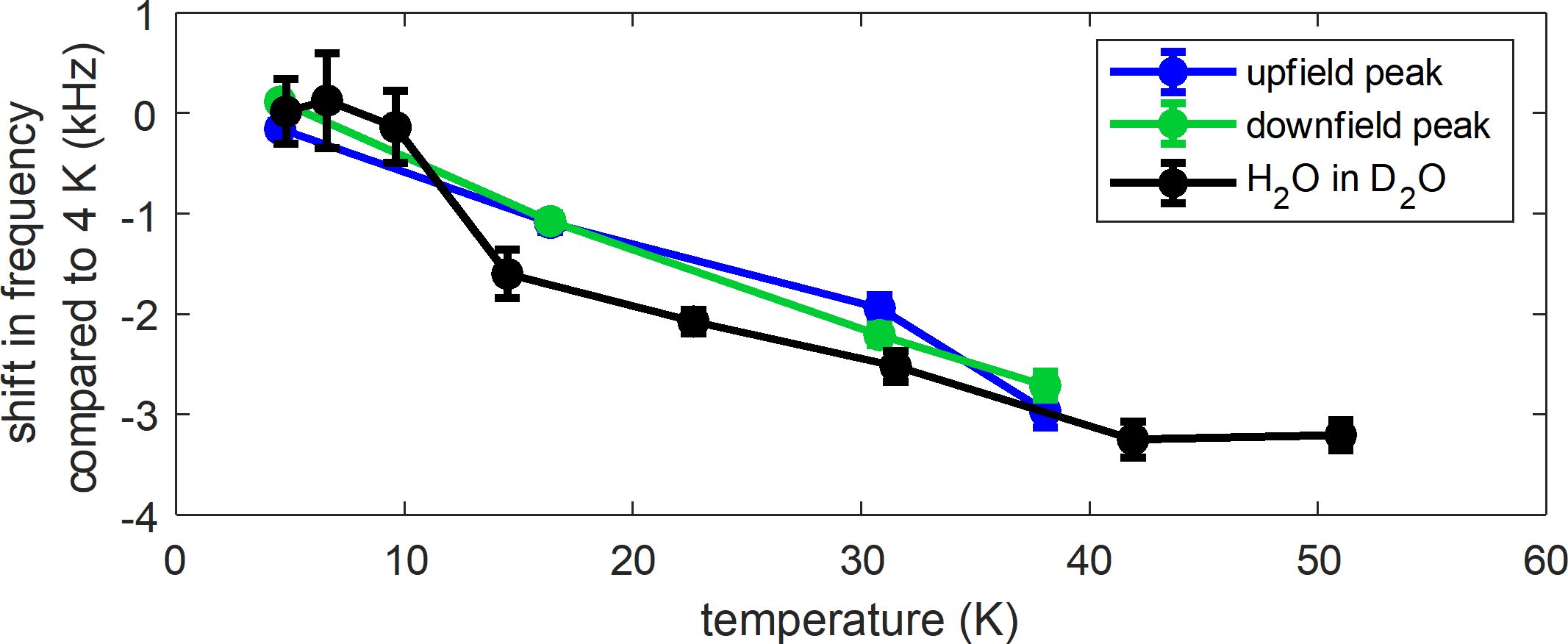}

\caption{Shift in frequency compared to the 4 K frequency of the two lines in the Si powder and of residual H$_{2}$O in D$_{2}$O measured as a function of temperature. The error bars
shown are given from the 95\% confidence interval of the integrated
intensity of each peak given from fitting of the NMR spectrum for
each point. \label{fig:tempdependence}}
\end{center}
\end{figure}

\noindent The Si powder spectrum was measured as a function of temperature from
40 K to 4 K in order to see if the large shift between the two $^{1}$H
resonances is temperature dependent. The temperature dependence of
the frequency of the two lines is shown in Figure \ref{fig:tempdependence}.  In the figure, the y-axis is the shift in frequency for each peak, compared to the peak's frequency at 4 K (i.e. $\omega(T)-\omega(4 K)$ for each peak). Also, in the figure, we plotted the shift in frequency of the $^{1}$H line of residual H$_{2}$O in D$_{2}$O. This line should not be temperature dependent, and therefore serves as a control experiment to help identify whether the peaks are actually temperature dependent or not.

As can clearly be seen, all three
curves have the exact same slope and therefore shift with temperature
in the same manner. Thus, the temperature dependence of the lines
is not a result of any changes in the NMR interactions but is rather
an artifact of shrinking of the probe or of the NMR coil at low temperatures. Thus, the frequency difference between the two peaks is also not temperature dependent in this temperature range. These results suggest that there is no temperature dependent Knight shift or paramagnetic shift in the Si powder line between 40 K and 4.2 K. 

\section{Frequency modulation}

\begin{figure}[H]
\begin{center}
\includegraphics[width=0.7\textwidth]{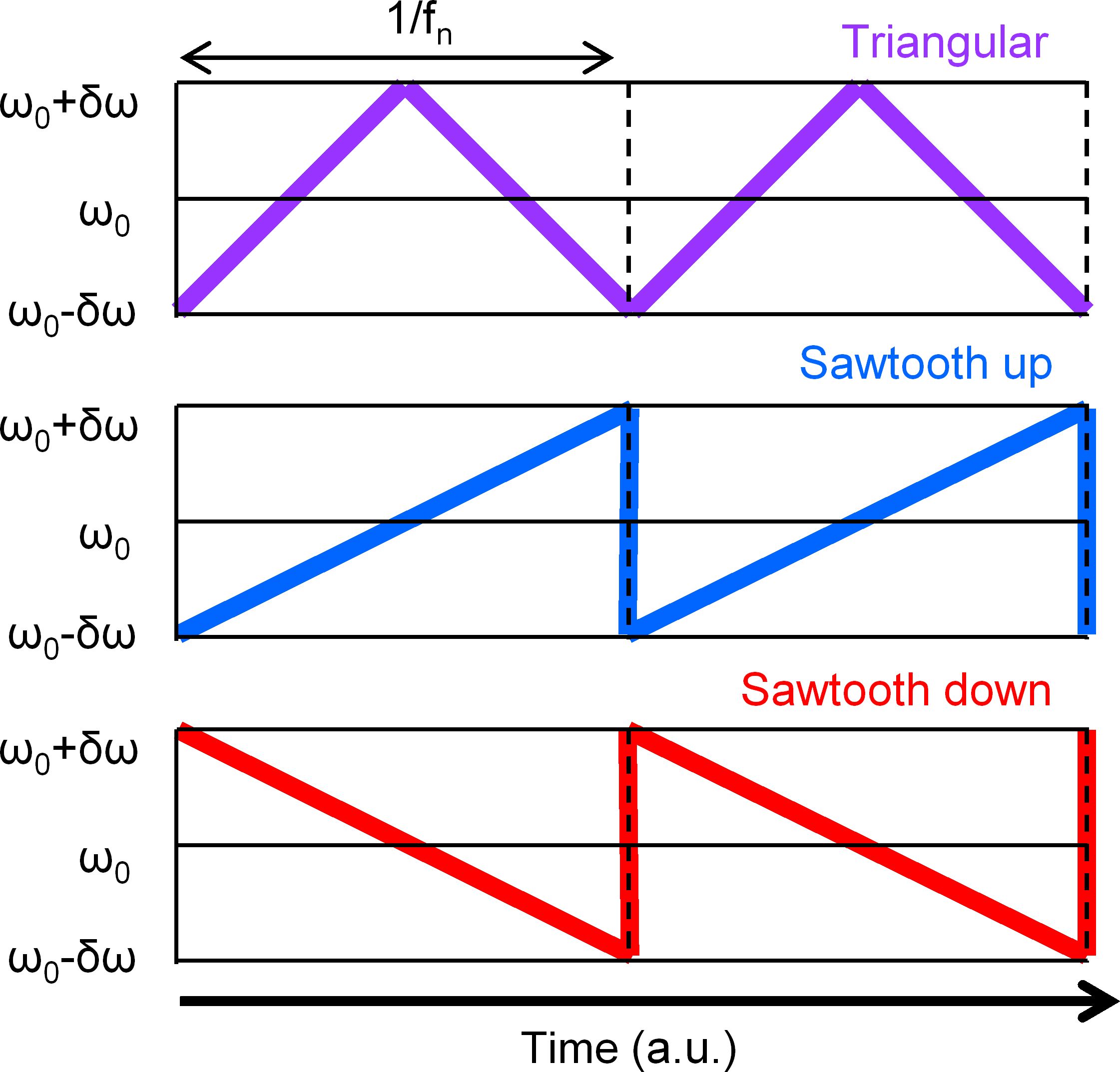}

\caption{Scheme showing the frequency modulation of the MW irradiation for triangular, sawtooth up, sawtooth down modulations.\label{fig:modulation_scheme}}
\end{center}
\end{figure}

\noindent DNP spectra were recorded with frequency modulation. The frequency
of the modulation (1/f$_{n}$ in Figure~\ref{fig:modulation_scheme}) refers to how many times per second
the frequency range was swept. The amplitude of the modulation, $\delta\omega$,
refers to the range of MW frequencies that was swept around the central
frequency, $\omega_{0}$. In all figures just the enhancement is plotted
at each central frequency. Three types of frequency modulation were
used: Sawtooth up, sawtooth down and symmetrical triangular modulation.
The frequency of the modulation was 10 kHz, and the amplitude was
70 MHz ($\pm$70 MHz around the central frequency) in all cases.

\section{Probe background}

\noindent We measured the probe background at 4 K using an echo with two different echo delay times: 30 $\mu$s and 200 $\mu$s. An empty glass tube wrapped in a small amount of teflon tape was placed inside the coil for these experiments. At 30 $\mu$s you can see that the background consists of a broad peak and a narrow peak. Lengthening the echo time to 200 $\mu$s effectively removes the broad peak from the spectrum, and makes the narrow peak much smaller. The narrow peak does overlap with the upfield peak and the downfield peak of our samples, however its intensity is negligible compared to the intensities of even the thermal spectra of our samples (where the teflon is only wrapped around parts of the sample that are outside of the NMR coil).

\begin{figure}[H]
\begin{center}
\includegraphics[width=0.7\textwidth]{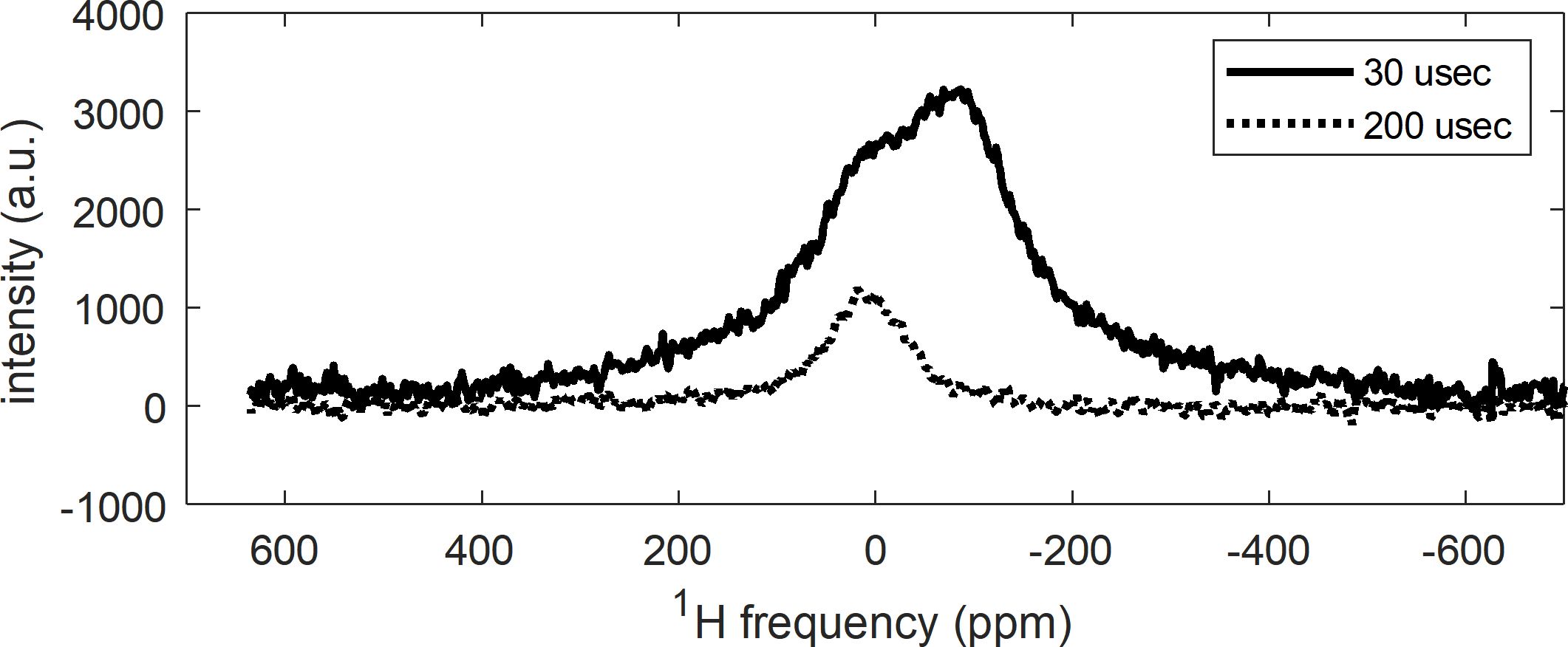}

\caption{Spectra showing the probe background measured at 4 K, with two echo times, 30 $\mu$s (solid line) and 200 $\mu$s (dashed line). Note the extended frequency range in the horizontal axis. \label{fig:background}}
\end{center}
\end{figure}

\section{NMR lineshape as a function of echo delay time}

\noindent We compare the $^1$H NMR spectra as a function of echo delay time, by plotting the spectrum measured with 30 $\mu$s and 200 $\mu$s. Both spectra were fit using the same two Lorentzian lines for the upfield peak and the downfield peak, varying only the peak intensity between them. The probe background is negligible with a 200 $\mu$s echo, as can be seen below. It is clear that even though 200 $\mu$s is a long echo time, it does not affect the NMR lineshape of our sample.

\begin{figure}[H]
\begin{center}
\includegraphics[width=0.7\textwidth]{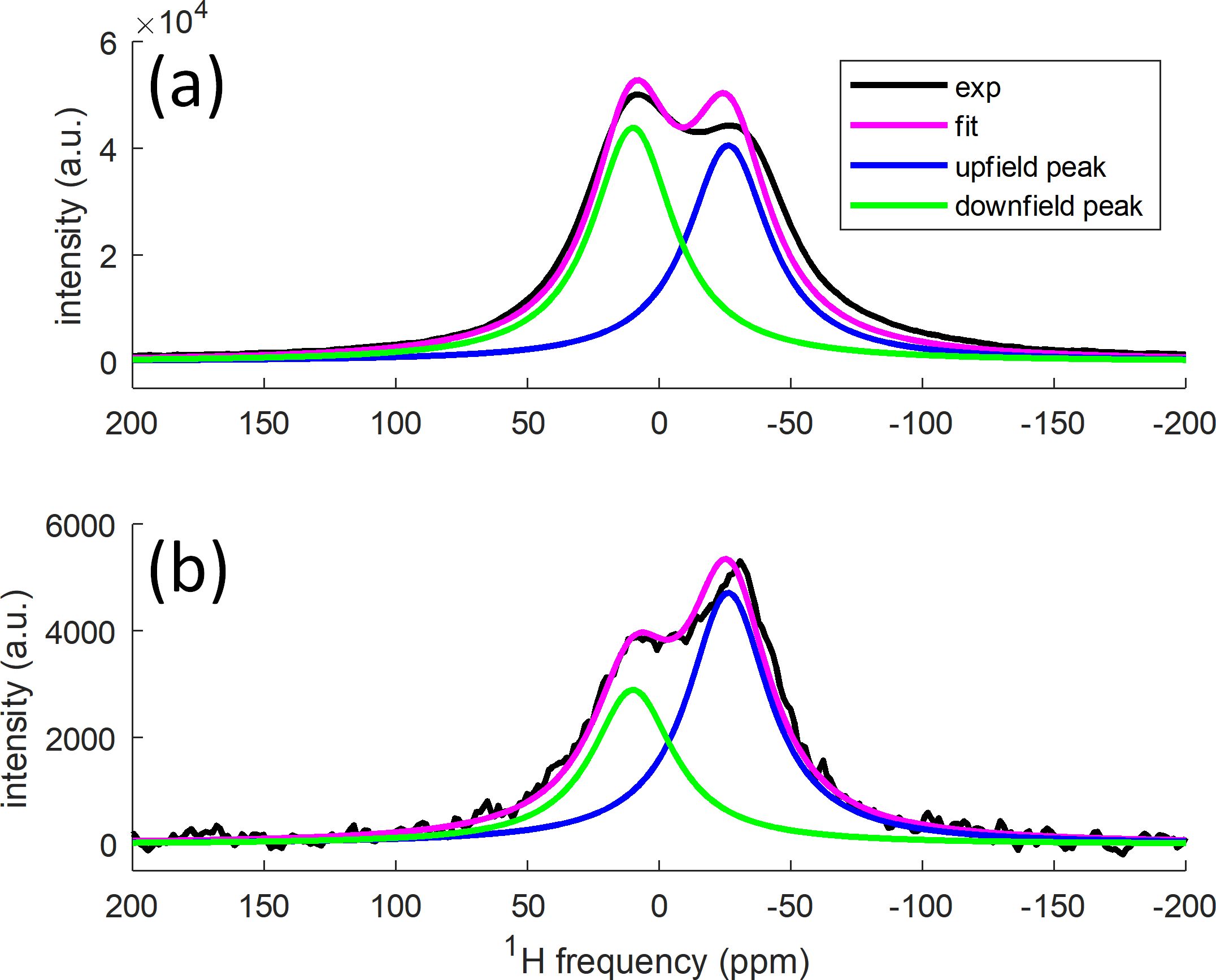}

\caption{Comparison of two $^1$H NMR spectra, with two echo times: 30 $\mu$s and 200 $\mu$s. Both spectra are fitted using the same two Lorentzian lines for the upfield peak and the downfield peak, varying only the peak intensity between them.  \label{fig:spectra_vs_echotime}}
\end{center}
\end{figure}

\section{Scanning electron microscope (SEM)}

\begin{figure}[H]
\begin{center}
\includegraphics[width=0.65\textwidth]{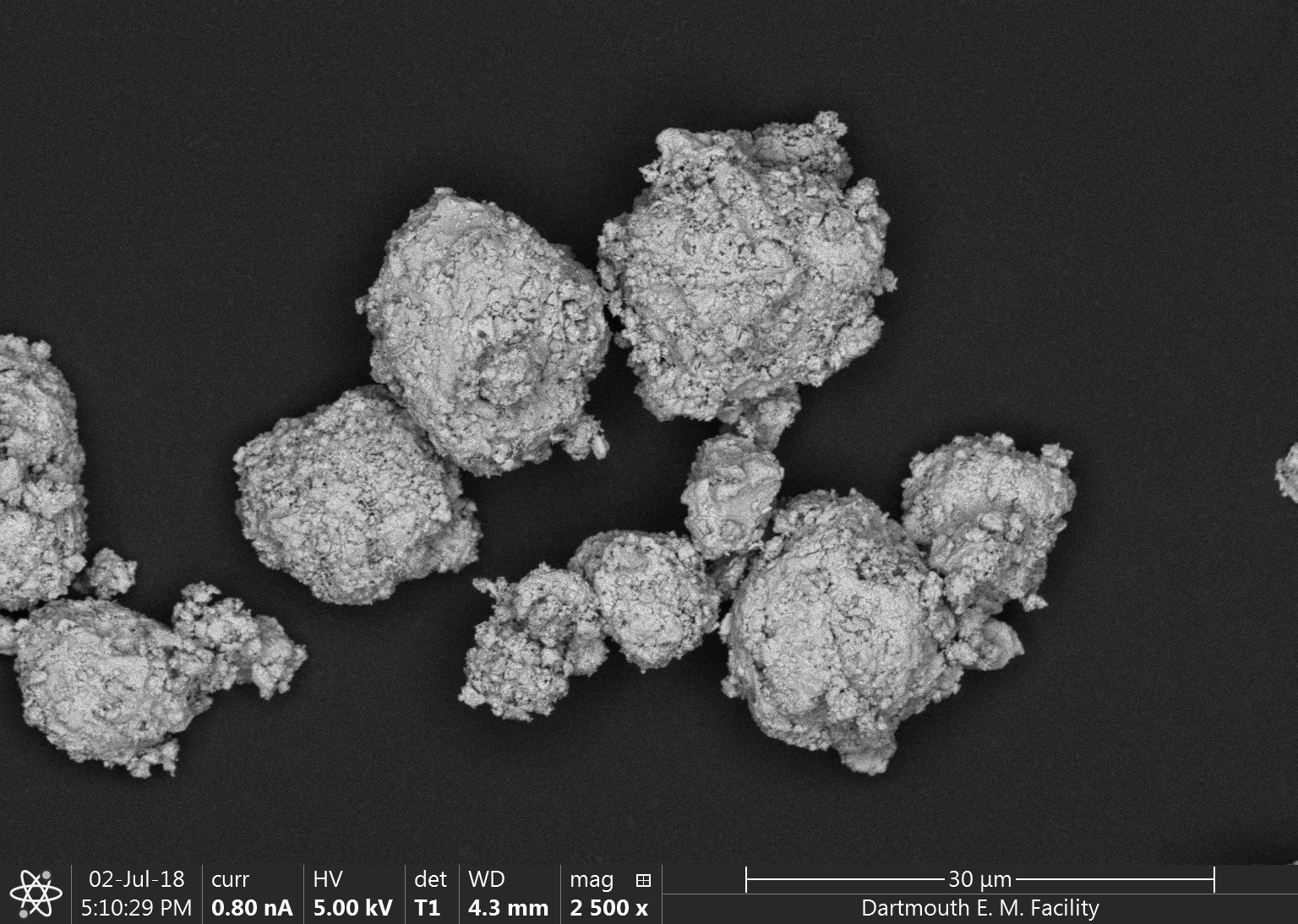}

\caption{SEM image showing the silicon particles.\label{fig:SEM}}
\end{center}
\end{figure}

\noindent The scanning electron microscope (SEM) image in Figure \ref{fig:SEM}
was taken on a FEI (Thermo Fisher Scientific) Scios2 LoVac dual beam
FEG/FIB SEM, using a T1 detector, an acceleration voltage of 5kV,
a current of 0.8 nA and a magnification of 2500 x. The silicon particles
were mounted to a specimen stage using double-sided tape. As can be
seen, particles reach up to 30 \textmu m in size, with the majority between  1 \textmu m and 15 \textmu m and are mostly spherical though they have rough edges which contain many small crevices and thus a large surface area.

\bibliography{references_Si_paper}